# Training Spiking Neural Networks for Cognitive Tasks: A Versatile Framework Compatible to Various Temporal Codes

**Chaofei Hong**
hongchf@tju.edu.cn
School of Electrical Engineering and Automation,
Tianjin University, Tianjin, P. R. China

## Abstract

Conventional modeling approaches have found limitations in matching the increasingly detailed neural network structures and dynamics recorded in experiments to the diverse brain functionalities. On another approach, studies have demonstrated to train spiking neural networks for simple functions using supervised learning. Here, we introduce a modified SpikeProp learning algorithm, which achieved better learning stability in different activity states. In addition, we show biological realistic features such as lateral connections and sparse activities can be included in the network. We demonstrate the versatility of this framework by implementing three well-known temporal codes for different types of cognitive tasks, which are MNIST digits recognition, spatial coordinate transformation, and motor sequence generation. Moreover, we find several characteristic features have evolved alongside the task training, such as selective activity, excitatory-inhibitory balance, and weak pair-wise correlation. The coincidence between the self-evolved and experimentally observed features indicates their importance on the brain functionality. Our results suggest a unified setting in which diverse cognitive computations and mechanisms can be studied.

## 1. Introduction

On the top-down viewpoint, the brain performs a large variety of functions ranging from perception and motion execution to higher cognitive processes such as reasoning and emotions; while viewing from the bottom, the diverse brain activities are mainly carried out by neurons transmitting and transforming spikes within the neural circuits[1]. How the brain organizes the spiking neural dynamics into meaningful computation has yet to be resolved. Nonetheless, most analysis and modeling of neural computation assume rate-based models, where the neuron represents information by the average firing rate[2]–[6]. Although rate-based approaches are straightforward, evidences have accumulated on the importance of exact spike timing for neural processing and network dynamics [7]–[10]. Accordingly, a number of temporal coding theories have been developed, such as synfire chains [11], polychronization [12], rank order coding [7] and predictive spike coding [13]. Yet, it remains challenging to confirm those theories by modeling studies due to the hardness of "handcrafting" complex cognitive functions into model networks.

Learning is the most essential element to achieve functionality for both biological and artificial neural networks. However, the endeavors of implementing biologically realistic learning rules on the functional spiking neural networks have only achieved limited success [14]–[18]. On the other hand, gradient-based learning on artificial neural networks (ANNs), especially with deep architectures, has achieved considerable success in various AI tasks [19]–[23]. Moreover, deep neural networks have begun to imitate cognitive functionalities such as memory and attention in solving tasks [24]–[26]. Despite those remarkable progresses, deep learning has rarely given significant feedbacks on how the brain works. This discrepancy, we believe, stems from two facts: first, the recent developments of deep learning were mainly guided by insights on mathematics, rather than neuroscientific findings; second, the highly simplified models in ANNs, which implicate rate-based computation, reflect little reality of the biological neural systems. Therefore, bridging the gap needs a unified approach, which are both compatible with the advanced tools from the deep learning and able to incorporate essential neuronal dynamics.

Spiking Neural Networks (SNNs) are widely used simulation model in the neuroscience field, which adopt biological models of neuron and synapses. Conventionally, the SNNs are constructed to model observed neural activities, and facilitate in explaining the underlying mechanisms[27]–[29]. The main obstacle of training functional SNNs is that the spike events make neuronal states incompatible with the standard, gradient-based learning methods (a notable exception is the recently proposed soft threshold model [30]). Some studies get around with this problem by regarding SNNs as pseudo rate-based models [31]–[34]. On another approach, SpikeProp considered the spike times as state variables and derived differentiable relationship between input and output spike times [35].The SpikeProp approach has been further developed focusing on the learning algorithms, and the SNNs have been shown equivalent to their rate-based counterparts in some simple tasks [36]–[43]. However, the efforts on SNN training are still far from solving real-world problems or replicating human-level cognitive functionalities. We find that two limitations are remained in current development: first, most of exist algorithms aim to learn exact spike times rather than various other temporal codes, which are potentially more suitable in certain tasks; and second, previous studies rarely explore the rich network structures and neural dynamics observed in biological neural networks that may facilitate neural computation.

In this paper, we adopt spiking neural network to model basic structural and dynamical characteristics of the neural system. A modified SpikeProp learning rule with improved the stability is introduced. The errors are assigned to each spikes (rather than neurons) through spike timing backpropagation. Conditions for efficient learning in SNNs are discussed and regulation methods are introduced to support a stable learning and neural activity. We introduced three temporal codes, which are suitable for different kinds of cognitive tasks. Then, we build a feedforward neural network to solve three simple cognitive tasks. In specific, we adopt rank-order code to learn digits recognition; we use the relative spike time code to transform point position in two coordinates; and we use the synaptic current as a code to generate motor sequences. In the three experiments, we explore different network dynamic modes from synchronization to sustained spiking activity. In addition, biological realistic characteristics such as lateral connections and sparse spikes are introduced into the model network, and several neural activity characteristics of the trained network have been investigated. By training and analyzing the three functional networks, we took a further step in relating the connectivity and activity to computational functionalities, and demonstrated the versatility of spiking neural networks.

## 2. Neuron Model and Learning Rule

### 2.1 Neuron Model

In this paper, we consider a leaky-integrate-and-fire (LIF) model with current-based synapses[44]. The dynamic of the neuron follows the differential equation:

$$v' = (s-v)/\tau_m \quad (1)$$

where $v$ is the membrane potential, $\tau_m$ is membrane time constant, and $s$ is the synaptic current. At each time the membrane potential reached the threshold $V_{threshold}$, a spike is evoked and the membrane potential is reset to $V_{reset}$. We model the synaptic current with a second-order differential equation:

$$\tau_r \tau_d s'' + (\tau_r + \tau_d) s' + s = \sum_k w_k \delta(t - t_k - d_k) \quad (2)$$

where the parameter $\tau_r$ characterizes the rise rate of the synaptic current and $\tau_d$ characterizes the decay, the Dirac $\delta$ function multiplied with synaptic weight $w_k$ relate the synaptic current to each spike $t_k$, and the parameter $d_k$ is the axonal conductance delay. Equation (2) indicated that the synaptic current is the linear convolution of incoming spike times with a prototype synaptic kernel. In addition, biological synapses can have diverse dynamics based on contained transmitters and receptors. For the first two experiments in this paper, we model the synapses with fast dynamics. In the third experiment, we considered synapses with a combination of fast dynamics and slow dynamics. Figure 1A shows four postsynaptic potentials (PSPs) induced by different synaptic receptor combinations. And a typical spike generation dynamics of the present model is depicted in Figure 1B.

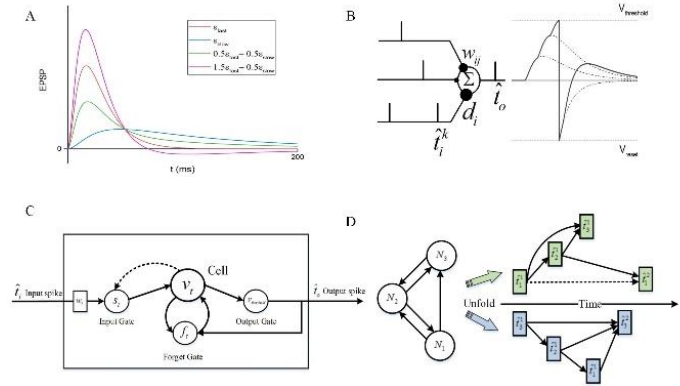

Figure 1. Description of spiking neural network model. (A) Postsynaptic potentials evoked by synapses with different dynamics. (B) Spike generation dynamics. Membrane potential reached threshold by linearly summation of postsynaptic potentials, and membrane potential is reset to Vreset. (C) Description of LIF neuron model under LSTM framework. (D) Different spiking order result in different computation graph of the neural network.

### 2.2 Learning Rule

Training spiking neural networks using gradient methods is first proposed in SpikeProp[35]. Following the same idea, we assume the errors are propagated to each neuron through the spikes. Because the output error is associated with given temporal code, we first decide the error propagation and updating rule and then discuss the cost function with different temporal codes. The network is assumed all-to-all connection to ensure the generality of the learning rule. Notably, network parameters such as thresholds and membrane time constants are also trainable using this learning rule. However, we only considered training for synaptic weights and axonal conduction delays, which can be updated using a fixed look-up table.

Given the partial derivative of cost function $E$ with respect to all spikes after the spike $t_i$. According to the chain rule, we can compute the partial derivative of $E$ with respect to $t_i$ as:

$$\frac{\partial E}{\partial t_i} = \sum_{j>i} \frac{\partial E}{\partial t_j} \frac{\partial t_j}{\partial t_i} + \left[\frac{\partial E}{\partial t_i}\right]_{out} \quad (3)$$

where $\frac{\partial t_j}{\partial t_i}$ characterizes the effect of spike $t_i$ to later spike $t_j$, and $\left[\frac{\partial E}{\partial t_i}\right]_{out}$ is an additional term for the output spike representing the part of error that defined by cost function. The errors for all the spikes can be obtained by iteratively apply equation (3) from the last spike to the first spike. After assigned the error to each spike, we can update the synaptic weights and conduction delays by:

$$\Delta w_{ij} = -\rho \frac{\partial E}{\partial w_{ij}} = -\rho \sum_{t_i^k \in N_i} \frac{\partial E}{\partial t_i^k} \frac{\partial t_i^k}{\partial w_{ij}} \quad (4)$$

$$\Delta d_{ij} = -\rho \frac{\partial E}{\partial d_{ij}} = \rho \sum_{t_i^k \in N_i} \sum_{t_j^n \in N_j} \frac{\partial E}{\partial t_i^k} \frac{\partial t_i^k}{\partial t_j^n} \quad (5)$$

where $\Delta w_{ij}$ is the change in synaptic weight $w_{ij}$, $\rho$ is the learning rate, $t_i^k \in N_i$ indicate all spikes neuron $i$ emit, $t_j^n \in N_j$ indicate all spikes neuron $j$ emit, and $\frac{\partial t_i^k}{\partial w_{ij}}$ characterizes the effect of weight $w_{ij}$ on spike $t_i^k$.

To complete the learning rule, we need to calculate $\frac{\partial t_j}{\partial t_i}$ and $\frac{\partial t_i^k}{\partial w_{ij}}$. Those two terms are dependent on the spike generation dynamics and therefore are neuron model specific. For ease of the derivation, we redescribe our model in the form of Spike Response Model (SRM)[1]. The neuron's membrane potential is given by:

$$v_j(t) = \sum_i w_{ij} \sum_f \varepsilon(t - t_i^f - d_{ij}) + \eta(t - t_j) \quad (6)$$

$$\varepsilon(t) = H(t) \left[\frac{\tau_m e^{-t/\tau_m}}{(\tau_m - \tau_d)(\tau_m - \tau_r)} + \frac{\tau_r e^{-t/\tau_r}}{(\tau_m - \tau_r)(\tau_d - \tau_r)} + \frac{\tau_d e^{-t/\tau_m}}{(\tau_m - \tau_d)(\tau_r - \tau_d)}\right]$$

$$\eta(t) = -V_{reset} H(t) e^{-t/\tau_m}$$

$$H(t) = \begin{cases} 0, & t < 0 \\ 1, & t \geq 0 \end{cases}$$

where $\varepsilon(\cdot)$ is the PSP kernel in response to input spike $t_i$, $d_{ij}$ is the conduction delay between neuron $i$ and $j$, and $\eta(\cdot)$ is the hyperpolarizing potential kernel which reset membrane potential to

$V_{reset}$ after a postsynaptic spike. Correspondingly, the partial derivative of $t_j$ in respect of each spike $t_i$ is:

$$\frac{\partial t_j}{\partial t_i} = \frac{\partial t_j}{\partial v_j}\frac{\partial v_j}{\partial t_i} \quad (6)$$

where $\frac{\partial t_j}{\partial v_j}$ characterizes the rise rate of membrane potential at the time of spike $t_j$, and $\frac{\partial v_j}{\partial t_i}$ characterizes the contribution of spike $t_i$ on the rise of membrane potential. If spike $t_i$ and $t_j$ are from different neuron, then equation (6) can be expend as:

$$\frac{\partial t_j}{\partial t_i} = \frac{w_{ij} d\varepsilon/dt\big|_{t=t_j-t_i-d_i}}{\sum_{t_k \notin N_j} w_{kj} d\varepsilon/dt\big|_{t=t_j-t_k-d_k} + \sum_{t_k \in N_j} d\eta/dt\big|_{t=t_j-t_k}} \quad (7)$$

Otherwise, if spike $t_i$ and $t_j$ are from the same neuron, then equation (6) can be expend as:

$$\frac{\partial t_j}{\partial t_i} = \frac{d\eta/dt\big|_{t=t_j-t_i}}{\sum_{t_k \notin N_j} w_{kj} d\varepsilon/dt\big|_{t=t_j-t_k-d_k} + \sum_{t_k \in N_j} d\eta/dt\big|_{t=t_j-t_k}} \quad (8)$$

And the partial derivative of $t_j$ in respect of each spike $w_{ij}$ is:

$$\frac{\partial t_j^k}{\partial w_{ij}} = \frac{\partial t_j}{\partial v_j}\frac{\partial v_j}{\partial w_{ij}} = \frac{\sum_{t_i \in N_i} \varepsilon(t_j - t_i - d_{ij})}{\sum_{t_k \notin N_j} w_{kj} d\varepsilon/dt\big|_{t=t_j-t_k-d_k} + \sum_{t_k \in N_j} d\eta/dt\big|_{t=t_j-t_k}} \quad (9)$$

### 2.3 Temporal Codes

A challenging issue for neural computation is to represent a variety of signals by the prototyped spikes. Various temporal codes have been proposed for different neural signals and different activity modes. In this section, we adopt three temporal codes in correspondent to three kind of tasks. The basic form of temporal code is to encode information by exact timing of all spikes. This temporal code utilizes the full representation capacity of the spikes and it is implemented by most SpikeProp kind algorithms. However, the requirement to access all spike times makes it hard for neural network to use and interpret. One way to alleviate this requirement is to narrow the time window of the temporal code. In fact, abundant experimental studies suggest that cortical neurons tend to spike coherently when attending to mental tasks [45]–[48]. Based on those experimental evident, an influential hypothesis named "Communication through Coherence" (CTC) proposes that neuron groups effectively communicate through gamma-band oscillatory synchronization [49]. In this scenario, each neuron can evoke at most one spike in the short time window.

Under the condition of CTC, we can use a relative time code to represent continuous variables by define a Linear-Nonlinear Model[50].

$$\tilde{t}_i = t_i - \bar{t}, \quad t_i = a\exp(-\frac{(\kappa_i \mathbf{x} - b)^2}{2c}) \quad (10)$$

Where $\tilde{t}_i$ is the relative time between spike $t_i$ and mean spike time $\bar{t}$, $\mathbf{x}$ denote the state vector the neural group encoded, $\kappa_i$ denote the linear filter represent receptive field of neuron $i$, the nonlinear function is Gaussian function with a scale factor $a$, bias $b$ and variance $c$, and the negative sign implies that neurons tend to spike early for strong stimulus. When a filtered stimulus $\kappa_i \mathbf{x}$ is sufficiently small, the neuron's spike time should approach to infinity and is equivalent to a failed spike. Hence, not all neurons are necessary to spike in this temporal code. Note that the exact spiking time $t_i$ is not directly observable, hence we need to decode state vector $\mathbf{x}$ through relative spike times $\tilde{t}_i$ using numerical methods. For simplicity, we can define the cost function as:

$$E = \frac{1}{2}\sum_i (\tilde{t}_i - t_i^*)^2 \quad (11)$$

where $t_i^*$ is the relative spike time represent the desired output. Correspondingly, the partial derivative of $E$ in respect of each spike $t_i$ is:

$$\frac{\partial E}{\partial t_i} = \tilde{t}_i - t_i^* \quad (12)$$

Another important realization of CTC is rank-order coding, which encode information by the order of the spikes while omit the spike timing[7]. This discrete temporal code is suitable for categorical variables where each spike order represents one class. We can define the cost function as:

$$E = \frac{1}{2}\sum_{i,j \in O} \Theta(\gamma_{ij}(t_i - t_j) + |\gamma_{ij}|\xi)^2 \quad (13)$$

$$\Theta(x) = \begin{cases} x, & x \geq 0 \\ 0, & x < 0 \end{cases}$$

where $O$ represent all the output spikes, $\xi$ represent a small time bin within which spikes are considered as spike coincidently, and $\gamma_{ij}$ characterizes the desired order between spike time $t_i^d$ and $t_j^d$: $\gamma_{ij} = -1$ for $t_i^d > t_j^d$; $\gamma_{ij} = 1$ for $t_i^d < t_j^d$; and $\gamma_{ij} = 0$ if $i = j$ or the order between $t_i^d$ and $t_j^d$ is not defined. Correspondingly, the partial derivative of $E$ in respect of each spike $t_i$ is:

$$\frac{\partial E}{\partial t_i} = \sum_{i,j \in \Gamma} \gamma_{ij} \Theta(\gamma_{ij}(t_i - t_j) + |\gamma_{ij}|\xi) \quad (14)$$

Above, we have introduced two temporal codes based on the assumption of CTC. However, the neural system often needs to generate signals that continuously evolve over time, such as motor sequences. Previous studies have proposed to use low pass filters to transform discrete spike trains into continuous time dependent variables[51]. Here, we define a kind of read-out neuron that only summate PSPs but do not evoke spikes. Then, we define a synaptic current code that uses the membrane potential as the network output:

$$x_j(t) = \sum_i w_{ij} \sum_f \varepsilon(t - t_i^f - d_{ij}) \quad (15)$$

Then we define the cost function as:

$$E = \frac{1}{2}\sum_j \int_D \left(x_j(t) - x_j^*(t)\right)^2 dt \quad (16)$$

Correspondingly, the partial derivative of $E$ in respect of each spike $t_i$ is:

$$\frac{\partial E}{\partial t_i} = \sum_j \int_D w_{ij} \varepsilon'\left(t - t_i - d_{ij}\right)\left(x_j(t) - x_j^*(t)\right) dt \quad (17)$$

Notably, both spike timing and spike order is not strictly constrained in this synaptic current code. Therefore, the timing of individual spikes is variable as long as the populational activity retains the same output current.

### 2.4 Remarks on Temporal coding and Spike Neural Network Computation

Applying gradient-based learning rule on spiking neural network implies that spike timing can represent information equivalently as the continuous variables in ANN. In addition, the neural dynamics has provided SNN several unique characteristics. Firstly, the gain of the inputs is not determined by static weights and activation function. As can be seen from the spike generation process in Figure 1B, the effectiveness of each input spike is dependent on the combination of synaptic dynamics and the dynamical state of postsynaptic neuron. Such property can be better interpreted by analog the spiking neuron to a long short-term memory unit (Figure 1C). In the input gate, the synaptic currents are determined by synaptic weights, and synapse dynamics; in the forget gate, the impact of each input is continuously being diminished by neuron membrane's leaky conductance; and for the output gate, the information of the inputs are transmitted by output spike only after the membrane potential reached threshold. Same as the original LSTM unit, spiking neuron model selectively transmit information based on its history inputs. To an extreme, the only information a spiking neuron processed is the temporal relation of the sparse all-or-none spikes. Meanwhile, the spiking neuron is much simpler than the LSTM unit in computational complexity. Secondly, the spiking neural network is a dynamical system and hence the computations are unfolded over time. Moreover, as mentioned above, the output spike selectively represents information in recent history inputs. Consequently, the structure of computation graph is dynamically determined by the spike timing, which is also the container of information (Figure 1D). As a result, training the spiking neural network with different inputs is equivalent to forge multiple 'effective' networks with shared weights.

### 2.5 Requirements for efficient learning and computation

Previous works on training the spiking neural network using gradient-based algorithms often report occasionally fails or slow in converge, especially for large data sets. The machine learning community has concluded some general requirements for efficiently training with gradient-based algorithms. Those requirements are rarely discussed in previous spiking neural network studies. Below, we will discuss two closely related conditions, which should also apply to learning in SNNs: 1. The mean and variance of neuron's states should be in a proper range. 2. The gradient of neuron's states should be stable through propagation.

In artificial neural networks, various normalization algorithms have been proposed to guarantee a proper mean and variance for the activation. The reason is to keep the activation efficient for gradient propagation and avoid "pathological curvature" and "covariate shift" [52], [53]. The activation of artificial neural network is analog to firing rate of biological neurons. For neural networks using temporal code, however, different definition is needed to descript the output states. Here we define the standard deviation of neuronal population's spike times as the variance of the output, and define the mean spike count as the mean of the output. Such definition is consistent with the definition in studies on synchronous spikes. Because the spiking neurons have a leaky nature in neural dynamics while they use spike time differences to represent information, the variance of spike times must fall in a proper range to transmit information efficiently. In addition, spiking neural networks also could suffer from "pathological curvature" and "covariate shift" problems during training for the same reason as artificial networks. Furthermore, because add or delete of spikes will change computation graph of the network. Hence, stable spike count during training is a fundamental requirement for efficient learning. Recently, a new activation function for ANN called SELU have been proposed, which can drive the mean and variance of the activation into a fixed attractor [54]. In neuroscience, many studies on synchrony spike propagation have acknowledged that spiking neurons also have similar properties [55], [56]. While those studies often focus on stable propagation of synfire chains, the spikes with wider dispersion can propagate stably as well (Figure 1E). However, the network's state is not guaranteed to be stable during training. As the connection weights are randomly initialized, the excitatory and inhibitory synapses are only loosely balanced. During training, however, the excitatory and inhibitory synapses learn to tightly canceling with each other to increase representation accuracy. Consequently, the firing rate of the network will decrease as the membrane potentials are driven more closely to rest potential. Hence, a regulation is needed to counter this fluctuation on firing rate.

Another important requirement for efficient ANN learning is stable gradients in propagation. If the gradients of errors vanish in propagation, the learning will be extremely slow in deep layers[57]. On the other hand, the learning will be unstable if the gradients explode in propagation[58]. Using a similar gradient-based rule, SNN learning should also suffer from those unstable gradient problems. Indeed, slow or unstable convergences are known issues for SpikeProp kind learning algorithms. Unlike learning in ANN, however, we find that the gradient explosion problem is more critical for SNN training. As the excitation-inhibition balance tends to become tighter and synapse weights tend to disperse during training, the gradients also tend to grow due to the intrinsic spike generation mechanism. Conventional methods to alleviate gradient explosion problem such as gradient clip can only solve half of the problem. Gradient explosion not only effects error backpropagation but also effects spike forward propagation, and it pushes the network into a chaotic realm where small noises would cause large changes in spikes (Figure 2A). Such chaotic networks cannot utilize spike timing to represent information (as argued by many neuroscientists who support rate coding), even if an ideal learning rule could fit the network to certain dataset. In addition, a stable propagation of spikes in SNN does not guarantee stable gradients, as is the case for ANNs (Figure 2B). To make matters worse, the gradient is very likely to explode when the membrane potential reaches threshold extremely slow (hair trigger condition). Although we do not find way to avoid gradient explosion completely, we reasonably alleviate the problem by regulate the raise rate of membrane potential. In biological neurons, adaptive currents can prohibit slowly ramping membrane potential from depolarization. Inspired from this fact, we set a minimum time derivative of membrane potential $\dot{V}_{threshold}$ as another spike threshold, which effectively avoids unreliable spikes.

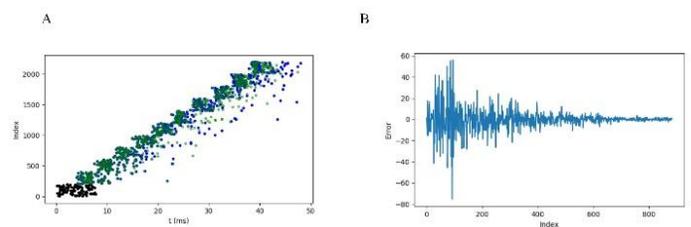

Figure 2. Example propagation of synchronous spikes and backpropagation of errors. (A) Synchronous spikes stably propagate through multiple layers, but a small disturb in input can cause large change of spikes in later layers under gradient explosion condition. (B) An example of gradient explosion of error backpropagation through spikes.

### 2.6 Weight Initialization and homeostatic regulation

In this section, we will describe the weight initialization and several regulation methods implemented in the network. We initialized excitatory and inhibitory synapse in different connections differently because of their different effects in spike initiation. For example, strong feedforward inhibitory synapses tend to cause silent neurons problem, while strong lateral excitatory synapses lead to avalanche phenomenon. Consequently, we initialize the feedforward connections with 80% of excitatory synapses drawn from positive half of the normal distribution $W_{exc} \sim N(0, \sigma_w)$, and 20% of inhibitory synapses with a fixed value $w_{inh}$ that roughly balanced with excitatory synapses. And for the lateral connections, the synapses are set with small fixed negative value $w_{lateral}$ in the synchronous spike mode; and they are initialized with 40% of excitatory synapses and 60% of inhibitory synapses in the last experiment to support sustained spikes. In addition, as the synapses cannot grow infinitely because of biological constrains, we add a weight-dependent factor to the original update rule:

$$\Delta w^* = \Delta w \cdot e^{\frac{-w \cdot sign(\Delta w)}{\tau_w}} \quad (18)$$

Where $\Delta w$ is the original weight update and $\tau_w$ is the constant control the limit of synapses weight. Furthermore, the stability of the network's firing rate is critical for efficient learning, but it can drifting away during training. Hence, we introduce a homeostatic rule on synaptic weights to regulate the firing rate:

$$\Delta w_i^h = (2r_{neuron} - r_i)\Theta(\frac{R_{layer} - R_j}{R_{layer}}) - r_i\Theta(\frac{R_j - R_{layer}}{R_{layer}}) \quad (19)$$

Where $\Delta w_i^h$ is the homeostatic regulation for synapse projected to neuron $i$, $r_{neuron}$ and $r_i$ are the desired and actual spike count of neuron $i$, $R_{layer}$ and $R_j$ are the desired and actual spike count of belonging layer of the neuron, and $\Theta(\cdot)$ is the ramp function.

## 3. Case studies

The neural system needs to perform various kinds of mental tasks, and neural assemblies are specialized to diverse functionalities. However, the cortical networks in different brain areas share a very similar microcircuit structure. This fact indicated that the neural network is capable to learn multiple tasks with a modular structure. In this section, we train a simple feedforward neural network to learn three typical cognitive tasks: classification task, coordinate transformation task, and movement generation task. The general network structure is shown in Figure 3A, where different layers are connected feed-forwardly. In addition, as the lateral connections are pervasive in cortical networks, neurons within the same layer are also inter-connected. All connections are initialized with a random conduction delay $d \sim \mathcal{U}(0,2)\ ms$, and the delays are constrained within 5 ms in the training. Other descriptions of the network are shown in Table 1. The network configuration has several simplifications for the ease of training: the network is only consist of homogeneous neurons; each neuron can have both excitatory and inhibitory synapses; and the network defined dense connections between layers. The simulation is based on the Brian simulator [59], with a simulation time step of 0.1 ms, and all the experiments use Adam as the optimizer [60]. The present three tasks only covered a small portion of all the mental tasks the brain performs. In addition, the simplified task description and network structure do not necessarily reflect the full biological reality. Nevertheless, our work provides an insight for the learning and computation capacity of spiking neural network.

### 3.1 Classification task

Creatures are constantly required to discriminate different objects or make appropriate decisions based on the sensory stimuli. Those tasks fall under the realm of classification problem in machine learning. To demonstrate the spiking neural network's capacity on those tasks, we trained a feedforward network to recognize handwriting digits in the MNIST dataset [61]. The MNIST dataset contains 60,000 labeled 28x28 grayscale images of handwriting digits for training and 10,000 labeled digits for testing. We use the relative time code to encode the image input, where each input neuron encoding the grayscale of one pixel. More efficient representation schemes can be achieved by designing sophisticated receptive fields of the encoding neuron using the linear filter $\kappa$, but it is unnecessary for this simple demonstration. A simplified version of rank-order code named time-to-first spike code is employed to encode classification result. We defined 10 output neurons, where each neuron is assigned a preferred digit class. The output is considered correct if the first spike neuron's preferred class matches the image label. The network consists of 310 LIF neurons (200-100-10 network) with the structure shown in Figure 3A and the neurons in relay layers are regulated to spike at average 0.5 spikes/trail. In addition, we add 10 kHz Poisson spikes to each neuron as background noise, and we trained a noiseless network as a comparison. We ran 20 epochs of training and a mini-batch size of 40 (Figure 3B). The training without noise is converge faster than with background noise, but training under background noise is more robust in test set (2.64% and 2.85% errors respectively).

Table 1  Network Model Specification

| Parameter | Value | Description |
|---|---|---|
| $\tau_m$ | 20 ms | Membrane time constant |
| $V_{threshold}$ | 1 | Spike threshold for potential |
| $\dot{V}_{threshold}$ | 0.025/ms | Spike threshold for raise rate |
| $V_{reset}$ | -50 | Spike resetting potential |
| $\tau_r$ | 2 ms | Raise time constant for fast synapse |
| $\tau_d$ | 8 ms | Decay time constant for fast synapse |
| $\rho$ | 0.02 | Learning rate for spike errors |
| $\beta_1$ | 0.9 | Adam optimizer's hyper-parameter |
| $\beta_2$ | 0.999 | Adam optimizer's hyper-parameter |
| $\varepsilon$ | 1e-8 | Adam optimizer's hyper-parameter |
| $\rho_{rate}$ | 0.04 | Learning rate for spike rate regulation |
| Additional specification for case study 1 | | |
| $w_{01}^{exc} / w_{01}^{inh}$ | 2.5 / 6.0 | Initial weight of input synapse |
| $w_{12}^{exc} / w_{12}^{inh}$ | 3.0 / 6.0 | Initial weight from relay layer 1 to 2 |
| $w_{11}^{inh} / w_{22}^{inh}$ | 0.1 / 0.1 | Initial weight of lateral connections |
| $w_{23}^{exc} / w_{23}^{inh}$ | 3.0 / 1.0 | Initial weight project to output layer |
| $w_{noise}^{exc} / w_{noise}^{inh}$ | 0.5 / 0.5 | Weight of random background activity |
| Additional specification for case study 2 | | |
| $a_{cor}$ | 8 ms | Scale factor of relative time code |
| $b_{posi}$ | 1.5 | Bias of position encoding |
| $c_{posi}$ | 1 | Variance of position encoding |
| $b_\theta$ | $0 \sim 2\pi$ | Preference rotation angle |
| $c_\theta$ | 2 | Variance of rotation encoding |
| $w_{01}^{exc} / w_{01}^{inh}$ | 1.5 / 3.6 | Initial weight of input synapse |
| $w_{12}^{exc} / w_{12}^{inh}$ | 1.0 / 2.0 | Initial weight from relay layer 1 to 2 |
| $w_{11}^{inh} / w_{22}^{inh}$ | 0.1 / 0.2 | Initial weight of lateral connections |
| $w_{23}^{exc} / w_{23}^{inh}$ | 3.6 / 2.0 | Initial weight project to output layer |
| Additional specification for case study 3 | | |
| $\tau_r^{slow}$ | 8 ms | Raise time constant for slow synapse |
| $\tau_d^{slow}$ | 100 ms | Decay time constant for slow synapse |
| $a_{motor}$ | 15 ms | Scale factor of motor command |
| $b_{motor}$ | $-1 \sim 1$ / $1 \sim 2$ | preference for motor command amplitude / frequency |
| $c_{motor}$ | 1 | Variance of motor command |
| $w_{01}^{exc} / w_{01}^{inh}$ | 1.5 / 3.0 | Initial weight of input synapse |
| $w_{12}^{exc} / w_{12}^{inh}$ | 1.5 / 3.0 | Initial weight from relay layer 1 to 2 |
| $w_{11}^{inh}$ | 0.1 | Initial weight of layer1 innerconnections |
| $w_{22}^{exc} / w_{22}^{inh}$ | 1.5 / 0.5 | Initial weight of layer2 innerconnections |
| $w_{out}$ | 0.2 | Initial weight project to output neuron |

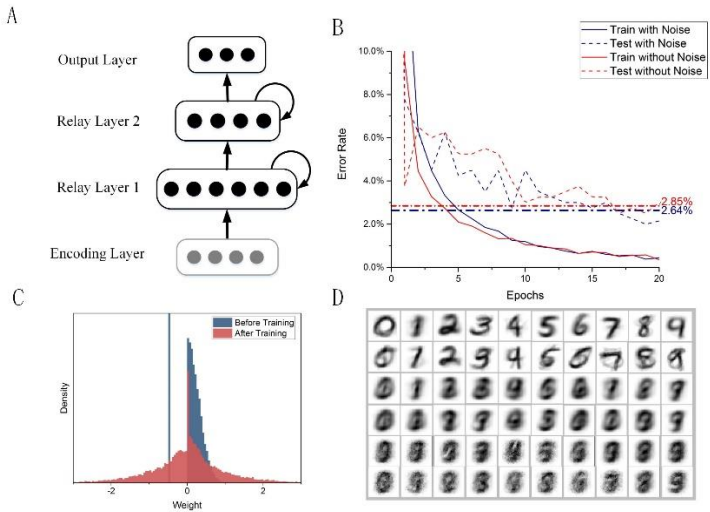

Figure 3. Training on MINST dataset. (A) Basic structure of the simple feedforward network. (B) Evolution of error rate during learning under noise and noiseless conditions. (C) Weight distribution before and after training. (D) Examples of learned features on MINST cognition task: features of layer2 (1) and layer1 (2) neurons estimated by first-spike triggered average (FSTA); features of layer2 (3) and layer1 (4) neurons estimated by spike triggered average (STA); and features of layer2 (5) and layer1 (6) neurons estimated by FSTA with random stimulus.

Next, we investigate the learned features of hidden layer neurons using spike triggered average (STA) kind methods. The STA is a conventional method in neuroscience to examine neuron's receptive field [50]. We conduct three different procedures to examine the receptive features: Firstly, we calculate each neuron's STA by applying the original digit images as the stimulus. Secondly, the first-spike triggered average for each neuron is calculated where we average the images that cause the neuron to spike first. Finally, we calculate the first-spike triggered average with a randomized stimulus, which each pixel is randomly chosen from pixels of the corresponding position in all images. As shown in Figure 3D, the neurons in the network have developed global features, where each feature characterizes a certain digit. The comparison of features calculated by STA and FSTA shows that the spike timing is more selective then spike count. In addition, we also notice that neurons response to randomized image stimuli with a much lower firing rate(~60 spikes) then those with digit images (~160 spikes). This phenomenon indicates that the neurons have been trained to exclusively response to certain features. To further confirm the effect of learning on network activity. We shuffled the postsynaptic weights within each neuron of the trained network, and stimulate the network with digit images and randomized images respectively. The shuffled network responds to both kinds of stimulus with about 250 spikes indistinctively. It shows that the network's activity can vary greatly with different detailed connections, even if the overall configuration remains the same. Such result questioned the validity of the conventional modeling approaches that reproduce neural activity observed in experiments with randomly generated connection weights.

### 3.2 Coordinate transformation task

Many cognitive tasks require the brain to represent outside world with internal models, which consist of properties that vary continuously. The predicting and transforming of those continuum properties in neural systems are analog to regression problems in machine learning. Coordinate transformation of object's position and velocity is a common task for both animals and robots. To demonstrate the SNN's capacity in regression problems, we train a feedforward network to perform a coordinate transformation task: we defined a 2-D coordinate system with 140 axes (7 origins and 20 directions each origin) and rotate the coordinate around the center with a certain angle (Figure 4A). The spatial organization of the coordinate is inspired from the coordinate system in hippocampus [62]. Different from the biological system, we use the relative time code to encode the position, with each coordinate axis assigned to one neuron as the linear filter. Besides, we defined 20 neurons to encode the rotation angle with a preferred angle uniformly range from 0 to $2\pi$. The network is consist of 800 neurons and 400 neurons for the first and second relay layers respectively, and those neurons are regulated to spike at average rate of 0.2 spikes/trail. We trained the network once with 64,000 randomly generated positions within a unit distance to the center and random rotation angles ranging from 0 to $2\pi$. At the end of the training, the mean spike time error reached 0.12 ms and the mean error of estimated position is 0.04. The transformation errors are not uniformly distributed in the coordinate space but tend to be larger in the outer areas (Figure 4B). This is most likely because that fewer neurons are dedicated to encode the outer areas.

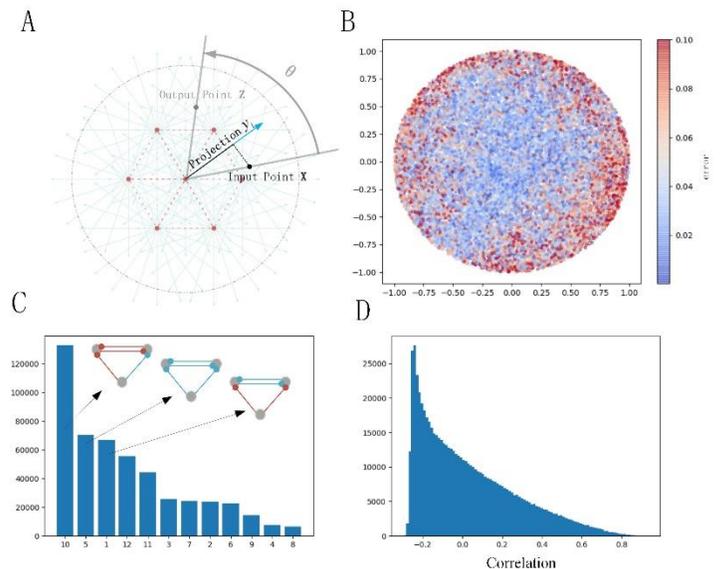

Figure 4. Training on corrdination transformation task. (A) Scheme of the coordiante and the coordiante transforamtion task: red point is the origins and light blue vector is the axises, the linear filtered position is caculated by project the point to each axises. The coordinate transformation is achieved by rotate the point with angle θ around the center. (B) The transformation errors on the output positions. (C) Unbalanced distribution of motif structures in the connections from layer 1 to layer 2. First three motifs are shown in graph: red connection represent positive weights, blue connections represent negative weights. (D) Distribution of pair-wise correlation between neurons within the same layers.

Next, we further investigate the structure and activity of the trained network. Experimental studies have shown that neural network structures are specialized such that the proportion of different motif structures is significantly deviated from chance level [63]. In the trained functional network, we also find that some motif structures consist a much higher portion than others (Figure 4C). Hence, those specific motif structures are highly likely to have some computational implications. Another commonly observed phenomenon in cortical network is the weak pair-wise correlation between globally synchronous spiking neurons [64]. Such phenomenon also arises spontaneously through training in the functional network even though it is not intentionally modelled (Figure 4D). The weak pair-wise correlation has been suggested to facilitate efficient coding, and our work has confirmed this hypothesis in a bottom-up approach. We also measured the receptive field of neurons in the relay layers, which shown organized patterns (Figure 5). The receptive fields of neurons in the two layers are quite similar. We have a closer inspection of the receptive fields by calculate the standard deviation of spike triggered input /output. It shows that layer 1 neurons are more sensitive to input position and rotation (std of input position: 0.66, rotation: 1.33 and output position: 0.85), while layer 2 neurons are more sensitive to output position (std of input position: 0.83, rotation: 1.43 and output position: 0.76).

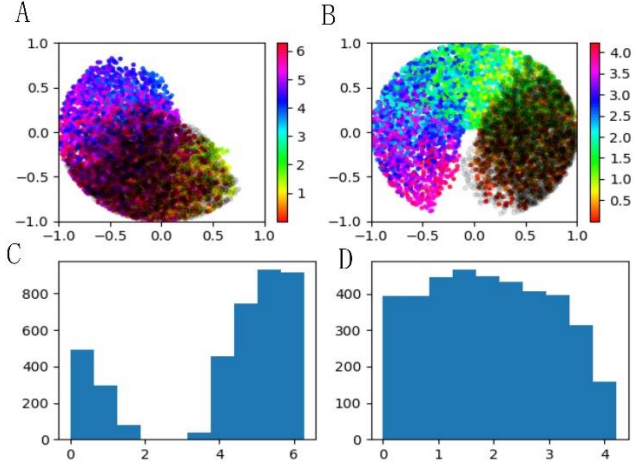

Figure 5. Developed feature selectivity of relay neurons in coordinate transformation task. (A) An example receptive field of neurons in relay layer1. Color represent rotation angle and dark area represent the transformed point coordination. (B) An example receptive field of neurons in relay layer2. (C) and (D) are distributions of rotation angles of stimulus responded by neurons in (A) and (B).

### 3.3 Motor sequence generation task

The brain is often required to process and transform signals between different time-scales. Those tasks often involve responding to particular temporal input sequences [65], responding after a delay [66], or responding with a temporally complex output[67]. To demonstrate the neural network's computational capacity in multi-time scale transformation tasks, we train the network to execute a series of motions according to the synchrony commands. The desired motion output is defined by a sinusoidal function: $f(t) = A\sin(\frac{\omega t L}{2\pi})$, where $A \in [-1,1]$ is the amplitude, $\omega \in [1,2]$ is the frequency and $L = 100$ ms is the duration of the motion. The experiment designed to last 200 ms each trail. And the output designed to maintain standstill in the first 100 ms and execute the motion output in the second 100 ms. We use the relative spike time to encode the input command with 100 neurons, where 50 neurons encode the amplitude and 50 neurons encode frequency with uniformly varied preferred stimulus defined by parameter $b$. The output is encoded by one synaptic current coding neuron receiving projections from the second relay layer. The synchronous commands are propagated through relay layer1 and transformed to the time-varying signal with sustained activity in relay layer2. Unlike previous configurations, the lateral connections in second relay layer are initialized with 40% excitatory synapses and 60% inhibitory synapses to support the sustained activity, and the output projections follows normal distribution $N(0, w_{out})$.

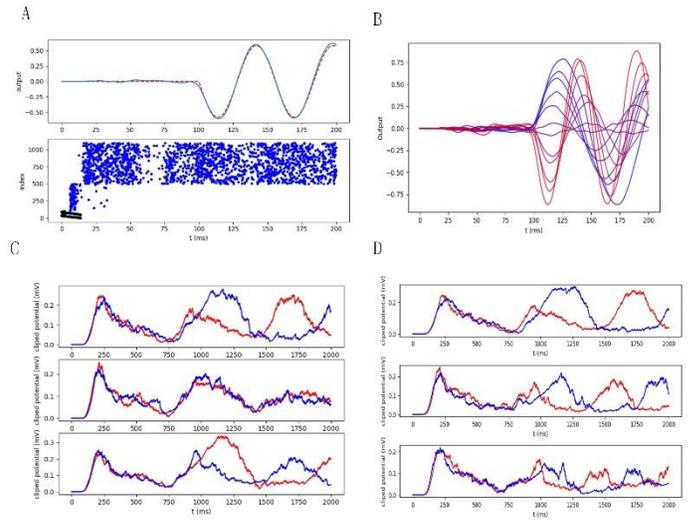

Figure 6. Trained network output in motor sequence generation task. (A) An example network activity and output motor sequence in response to the motor command. (B) A series of motor sequences generated by the network. the amplitude is uniformly varied from 1 to -1 and the frequency is uniformly varied from 1 to 2. (C) The average positive membrane potential of two group of relay layer 2 neurons. The red represent excitatory projection neurons and blue represent inhibitory projection neurons. The amplitude of the output is varied from 1 (1) to 0 (2) and -1 (3). (D) Same as (C), but with fixed amplitude and the frequency of the output is varied from 1 (1) to 1.5 (2) and 2 (3) cycles/100ms.

We trained the network through 3000 iterates with 28 random samples each mini-batch, and the typical execution is shown in Figure 5A. If we only update the weights between the second relay layer and the output neuron, then the network is an implement of Liquid State Machine (LSM) [68]. However, we find updating weights in deeper layers increased both the speed and capacity of learning. In addition, the synapse in this network contains a fast and a slow dynamics characterized by two separate weights. The slow synaptic dynamics provide a long-term channel to propagate errors from large time-scale signal to short time-scale spike, while the fast synaptic dynamics manage the transient spike dynamics by correct the residual errors. Notably, another recently proposed method also used synapses with fast and slow dynamics to learn motions [51]. Differ from our work, their method uses fast synaptic dynamics as a reference and learn the motion by update slow synaptic weights. The aims of those two works are also different: [51] aims to transform one certain input time-varying signal to one certain output in the same time-scale, while our work aims to transform a set of synchrony commands into a series of motion outputs (Figure 5B). To further investigate the network dynamics under different motion execution, we measured the mean positive membrane potential of two groups of relay layer2 neurons, which project excitatory and inhibitory outputs respectively. In Figure 5C, we fixed frequency of the motion and varied amplitude from 1 to 0 and -1. It shows that the relative phase of two neuron group activities changed accordingly. In Figure 5D, we fixed the amplitude of the motion and changed the frequency from 1 to 1.5 and 2. As expected, the oscillation frequency of neurons also followed the frequency of the desired motion. Similar to the first experiment, this result have also demonstrated that the overall neural dynamics can be precisely determined by exact spike times.

## 4. Discussion

Spiking neural networks have been a useful tool for neuroscience in modeling neural dynamics. Using the gradient-based rule, we demonstrated that the SNNs are capable of learning different functions with different temporal codes. It has been proposed recently that synaptic plasticity mechanisms may have achieved some form of error back-propagation [69], [70]. Admittedly, the exact relation between synaptic plasticity and gradient back-propagation still needs further study. It is safe to say, however, that the trained spiking neural network can provide valuable insights on neural coding and neural computation regardless of the learning rule. The three tasks we demonstrated here are very simple, but they covered the functionalities of classification, internal representation, and long short time-scale transformation, which are essential for other complex mental tasks. Our neural network model is considerably simplified and do not well reflect the organization of real cortical networks. Nevertheless, many important features of cortical network have evolved during the training, such as specialized receptive fields [71] and motif structures [63], tight excitation-inhibition balance [72], and weak pair-wise correlation in spikes [64]. The spontaneous development of those features suggests that they may be necessary for the functionality of neural networks. Hence, building functional network through training may provide a new way to investigate the role of the network features on brain functions. In addition, we found in the experiments that the overall activities of the trained network diverge greatly in response to meaningful and random stimulus. The selective activities, such as selective synchronization among brain areas, are typical properties in brain. Our work suggested that those phenomena might result from the specific connectivity learned in specific tasks rather than the special organization in the brain. Therefore, training or biological learning could also be an indispensable step on modeling neural behaviors. Additionally, it has been proposed recently that the recurrent neural networks (RNNs) could be versatile tools of neuroscience research [73], [74]. The trained RNNs and SNNs can be complementary to each other: the RNNs are more efficient in describing large-scale activities while SNNs can describe more details such as transient dynamics and exact spike activities. Altogether, training can provide a powerful enhancement for the current modeling study on neural computation and cognitive functions.

Alongside training the three tasks, we have spotted three important requirements for efficient SNN learning, and we modified the network model and learning rule accordingly. The first requirement is proper spike time variance: On one hand, spike times need to contain enough entropy to represent rich information with temporal codes. On the other hand, relevant neurons need to spike within an effective time window confined by the postsynaptic spike generation dynamics. With a proper weight initialization and regulation, we have shown that the spike time variance can be self-organized within a proper range in SNNs. The second requirement is proper and stable spiking rate: Even though our network encodes information by temporal codes, stable spikes are still necessary for effective information transmission and stable learning. On the other hand, it also should be flexible enough to allow irrelevant neurons cease to spike at certain stimulus conditions. In addition, the spike sparseness is also preferable for nonlinear neural computations [75], [76]. Hence, we introduced a spiking rate regulation method to keep the neurons at a low and stable firing rate. The third requirement is stable gradient propagation: Gradient vanishing would cause inefficiency in learning, while gradient explosion would cause instability of both learning and network activity. The gradient vanishing problem is benign in SNNs with lateral connections, and we introduced a threshold for the rise rate of membrane potential to avoid gradient explosion caused by unreliable spikes. However, stable gradients during training are still not fully guaranteed in our model, and further studies are needed to solve this problem. Notably, our experiments are based on SpikeProp kind learning using LIF neuron networks. The linear additive requirement of the gradient-based rule posed a strong constrain on further development. On another thread of studies, local error assignment rules, which do not need exact gradients, have been developed such as difference target propagation[77], random back-propagation [78], and synthetic gradients [79]. Those studies are currently focused on ANNs and rate-code based SNNs. However, the rate code is consist with codes based-on relative spike times, since a neuron with a high firing rate is expected to spike early[80]. Hence, we expect to extend those rules to networks with transient spiking dynamics in further studies.

The fields of neuroscience and artificial intelligence have a long and intertwined history. Recently, researchers in the AI field have called for more collaborations between the two fields [81], [82]. One main aim of this study is to demonstrate that SNN can be a common language bridging neuroscience and deep learning. Thus, on one hand, we can introduce advance technics in deep learning to train biological neural network models. On the other hand, the SNN also have its own intrinsic properties, which can extend the capability of machine learning systems. On the remark of SNN computation, we have likened the LIF neuron model to the LSTM cell. Notably, the LIF model is an extremely simplified version of spiking neuron. More complicate biological neuron models may have different realizations for the gates in the LSTM analogy. For example, the input gate is also controlled by membrane potential in models with conductance-based synapses; and for the recently proposed soft threshold model [30], it allows the output to change gradually through a voltage-dependent gate function. An essential difference between SNNs and RNNs, however, is that the dynamical evolvements of SNNs are mostly constrained within the neuron cell, while the dynamics in RNNs are defined as a network property. The additional single neuron computation renders spiking neural network the ability to dynamically routing information flow by spiking activities [83].This property is applicable both on the scale of microcircuit and large neural assemblies. As only a very small portion of neurons is active at any given time in the brain, very specific communication can be established through these synchronizations [49]. The ability to flexibly routing the information flow in the brain is considered fundamental for brain's multitask capability [83], [84]. In this work, we introduced a synapse model with both fast and slow dynamics, which transform a spike signal into two different time-scales. In further studies, we can introduce heterogeneous intrinsic neuronal properties such as adaptive currents and dendritic nonlinearities[85], which will project spike timing information to higher order temporal spaces. As we have shown in the experiments, the spiking activity can support different function modules with different temporal codes. Combined with the flexibility in information routing, we believe that the large-scale spiking neural networks with biological network dynamics and diverse temporal codes can be a good candidate for general AI systems. Furthermore, the SNN configuration can be more energy efficient in engineering perspective. Rather than communicate globally in each update cycle, the spiking neurons only need to communicate to others if the internal computation evoked a meaningful event. Adopting such event-driven computing architecture, neuromorphic computer chips such as IBM TrueNorth [86] and SpiNNaker [87] can simulate millions of neurons in real time with relatively low energy assumption. Such properties can make SNNs more preferable in real life applications.